\documentclass{elsart}

\usepackage{graphicx}

\usepackage{amssymb}

\begin{document}

\begin{frontmatter}



\title{Solar gravitational energy and luminosity variations}


\author[grasse,tabriz]{Z. Fazel}
\author[grasse]{J.P. Rozelot}
\author[paris]{S. Lefebvre}
\author[tabriz]{A. Ajabshirizadeh}
and \author[nice]{S. Pireaux}
\address[grasse]{Observatoire
de la C\^ote d'Azur, GEMINI Dpt., and UNSA University (Fizeau Dpt.), Av. Copernic, 06130
Grasse, France, email to: nayyer.fazel@obs-azur.fr and jean-pierre.rozelot@obs-azur.fr}
\address[tabriz]{University of Tabriz, Faculty of Physics, Dept. of Theoretical Physics
and Astrophysics, Tabriz, Iran, email to: a-adjab@tabrizu.ac.ir}
\address[paris]{Laboratoire AIM, CEA/DSM, CNRS, Universit\'{e} Paris Diderot, DAPNIA/SAp, 91191 Gif sur Yvette cedex, France, email to:sandrine.lefebvre@cea.fr}
\address[nice]{Previously at: Observatoire de la C\^ote d'Azur, ARTEMIS Dpt., Av. Copernic, 06130
Grasse, France; Now at: Observatoire Royal de Belgique, Dpt. 1, 3 Av. Circulaire, 1180 Bruxelles, Belgique, email to: sophie.pireaux@oma.be}
\begin{abstract}
Due to non-homogeneous mass distribution and non-uniform velocity rate inside the Sun, the solar outer shape is distorted in latitude.
In this paper, we analyze the consequences of a temporal change in this figure on the luminosity.
To do so, we use the Total Solar Irradiance (TSI) as an indicator of luminosity. Considering that most of
the authors have explained the largest part of the TSI modulation with
magnetic network (spots and faculae) but not the whole, we could
set constraints on radius and effective temperature variations.
Our best fit of modelled to observed irradiance gives $dT$ = 1.2
$K$ at $dR$ = 10 mas.
\newline
However computations show that the amplitude of solar irradiance modulation is very sensitive to photospheric
temperature variations. In order to understand discrepancies
between our best fit and recent observations of Livingston et al.
(2005), showing no effective surface temperature variation during
the solar cycle, we investigated small effective temperature
variation in irradiance modeling. We emphasized a phase-shift
(correlated or anticorrelated radius and irradiance variations) in
the ($dR$, $dT$)--parameter plane.
\newline
We further obtained an upper limit on the amplitude of cyclic solar radius variations between
3.87 and 5.83 km,
deduced from the gravitational energy variations. 
Our estimate is consistent with both observations of the helioseismic radius through the analysis of $f$-mode frequencies and observations of the basal photospheric temperature at Kitt Peak.
\newline
Finally, we suggest a mechanism to explain faint changes in the solar shape
due to variation of magnetic pressure which modifies the granules
size. This mechanism is supported by an estimate of the
asphericity-luminosity parameter, {\bf {\textit
w }} = -7.61 $10^{-3}$,
which implies an effectiveness of convective heat transfer only in very
outer layers of the Sun.

\end{abstract}


\end{frontmatter}

\section{Introduction}

If to first order the Sun may be considered as a perfect sphere, it is clear that due to its axial rotation,
the final outer shape
will be a spheroid. Moreover, the distribution of the rotation velocity being far from uniform both at the surface and in depth,
this final figure will be more complex.
Although the resulting asphericities are very small, some open questions
which remain are: to know if the passage from a sphere to a
distorted shape will affect the luminosity, and if so, to quantify
this effect. The first point has been partially studied in
Rozelot $\&$ Lefebvre (2003) and in  Rozelot et al. (2004). The second point was first addressed in Fazel et al.
(2005) or Lefebvre et al. (2005). The present paper shows how irradiance and temperature observations allow us to put strong upper limits on radius variations.
We use the TSI as an indicator of solar luminosity. Indeed as luminosity changes, so does the basic level of the
TSI, which is additionally modulated by surface magnetic activity
(spots, faculae, and network). This is not a minor question as the
TSI variation is often claimed to be of magnetic origin alone. Mechanisms
which may produce changes in irradiance have been discussed since
years, but we are still unable to propose a full comprehensive
model. As pointed out by Kuhn (2004), two different processes are
proposed. One involves surface effects (see for instance Krivova et al. 2003), 
and the other is due to a complex heat
transport function from the tachocline to the surface, including
global properties, mainly magnetic field, temperature and radius
(Sofia, 2004). Models based on the assumption that the
irradiance variations on time-scales longer than a day are entirely
and uniquely caused by changes in surface magnetism are rather successful (Krivova and Solanki, 2005), as
correlative functions between observed and modelled data show an agreement of 
{\large $_{\verb ~ } $}
90-94 \%.
However, the main observations which have not yet been reproduced by these models are
brightness changes measured by limb photometry (Kuhn et al.,
1988; Kuhn and Libbrecht, 1991). Furthermore, the recent SoHO/MDI experiment
has proved that exceedingly small solar shape fluctuations are
measurable from outside our atmosphere (Emilio et al., 2007). Accordingly, efforts should
be made to use these additional observations to better constrain
solar model parameters (radius, temperature) and
possibly the proportion of irradiance changes produced by
surface magnetism. We think that there is still room for improvements.
This paper is an attempt to clarify if some
of the $6-10\%$ of total solar irradiance left unmodelled by surface magnetism could be of other origin: from this point of view the
variability of the global distorted shape of the Sun must be explored.

In the following section, we will show how variations of the distorted outer
shape of the Sun contribute to a fraction of TSI variations,
assuming the main part of TSI variations being modelled by
magnetic mechanisms. We will also emphasize the key role of
surface effective temperature.
\newline
In Section 3, we will illustrate the lack of consensus between present observations of
solar radius variations (apparent radius) from the point of view
of amplitude and phase with respect to the solar cycle.
Moreover, there exist discrepancies between observations and
theoretical models regarding such variations. Hence new
observations (especially space--dedicated missions) are needed.
\newline
In Section 4, we will explain how variations of the
gravitational energy in the upper layers of the convective zone
may imply solar radius variations. According to the observed
amplitude of irradiance variations, we set an upper bound, of a
few kilometers only, on solar shape changes. This last model shows that solar radius variations are anticorrelated with irradiance variations during the solar cycle. We will then provide additional
information on the localization of luminosity variations by
computing the asphericity-luminosity parameter ({\bf {\textit w}}).
\newline
In Section 5, we suggest a mechanism to describe the connection between
solar radius and magnetic activity.
\newline
Finally, in Section 6, we present our conclusions.

\section{ Solar radius variations and luminosity changes  }

The ``outer shape" of the Sun must be defined: the Sun has an extended atmosphere and it is not so simple to determine the upper
limit of its photosphere. One of the most simple approach is to define this shape as an equipotential surface with respect to
the total potential (gravitational and rotational).
But, a contrario, if this definition has a physical meaning, the method to measure the true radius of the Sun, whether from space
or from the ground, is unclear. The observed solar radius, which is apparent, may be different from the theoretical radius,
whatever the definition of the latter is (see section \ref{solarradius}). Moreover, it is expected from the above definition
that the radius, $R$, is a function of latitude ($\theta$), both from an observational point of view
(Rozelot et al. 2003, Lefebvre et al. 2004) and a theoretical one (Armstrong and Kuhn, 1999, Lefebvre and Rozelot, 2004).
That is, at a constant pressure $p$ :
\begin{equation}
R(\theta){\mid}_{p} = R_{sp} \left[1 + \sum_{n,~even} c_{n} P_n(\theta)   \right]
    \label{rayonvecteur}
\end{equation}
\noindent
where $R_{sp}$ is the radius of the best sphere fitting both polar ($R_{pol}$) and equatorial ($R_{eq}$) radii $\left( = \sqrt [3] {R_{eq}^2 R_{pol}}~ \right)$, $c_{n}$ are
the shape coefficients (related to ``asphericities") and $P_n(\theta)$ are the Legendre polynomials of degree $n$ ($n$ being
even due to axial-symmetry).
We need to compute the solar surface area $A$, corresponding to Eq. \ref {rayonvecteur}:
\begin{equation}
  A = 4\pi \int^{\pi/2}_{0} R(\theta) \left[1+\left(\frac{dR \,(\theta)}{d\theta}\right)^2\right]^\frac{1}{2} d \theta.
  \label{surface}
\end{equation}
\noindent Armstrong and Kuhn (1999) or Rozelot et al. (2004)
provided estimates of the shape coefficients. The best available
values are $c_2$ $\in$ [$-2$$\times$$10^{-6}$,
$-1$$\times$$10^{-5}$] and $c_4$ $\in$ [$6$$\times$$10^{-7}$,
$1$$\times$$10^{-6}$]. For convenience, we express these results
in fractional parts of the best sphere $A_{sp}$ = 6.087 $\times$$10^{+18}$ $m^{2}$  which corresponds to the radius
$R_{sp}$= 6.959892$\times$$10^{8}$
$m$. Computations were carried up to $n$ = 4, leading to
$dA$($c_2$,$c_4$)/$A_{sp}$ $\in$ [$1.82$$\times$$10^{-6}$,
$6.37$$\times$$10^{-6}$], where the minimum corresponds to the
lower bound of $c_2$ and $c_4$ given above, while the maximum
corresponds to their upper bound. Those values can be compared to
the ones deduced from an ellipsoid\footnote{The area of an
ellipsoid of radii $R_{eq}$ = $a$, $R_{pol}$~= $b$ and
$c$=$\sqrt{a^2-b^2}$ is given by:
\begin{equation}
 A_{ell} = 2 \pi \left[a^2 + (a b^2 /c) \ln \frac{a + c}{b} \right ]
    \label{aireellipsoide}
\end{equation}
} of radii $R_{eq}$ = $a$ and $R_{pol}$ = $b$ with $R_{eq}$ = 6.959918$\times$$10^{8}$ $m$ and $R_{pol}$ = 6.959844$\times$$10^{8}$ $m$,
when $dR$ (= $da$ = $db$) 
varies from 10 mas
to 200 mas
(the choice of these two values will be explained later; see also Rozelot and Lefebvre 2003): $dA/A_{ell}$ $\in$ [$3.08$$\times$$10^{-6}$, $6.16$$\times$$10^{-5}$].

Let us call $F_{r}$, the radial component of the energy flux vector \textbf{F}. In the two-dimensional case, the luminosity, $L$, depends on $\theta$ :
\begin{equation}
L = 2 \pi \int^{0}_{\pi} r^2 F_{r}(r, \theta, t) \sin\theta d(\theta)
    \label{lumi}
\end{equation}

\begin{figure}[t]
\begin{center}
\includegraphics[width=12.3cm,height=6.4cm]{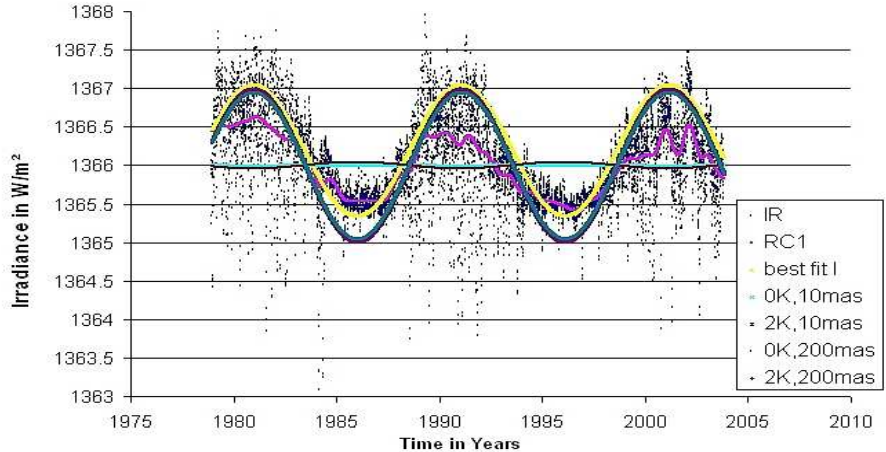}
\end{center}
\caption[]{Total irradiance variations with time. This figure shows the observed composite irradiance versus time (called IR, dots),
according to dataset updated to 01/10/2003 (Fr\"ohlich and Lean, 1998); the first component {\emph {RC1}} in the Singular
Spectrum Analysis (trend); the best sinusoidal curve fit to the observed composite data with $P$= 10.09 yrs and $\phi$= 1.026 rad;
and four sinusoidal models with different appropriate pairs of [$T$ (in K), $R$ (in mas)], as indicated in the right box. }
\label{irradiancessa}
\end{figure}

We start from the suggestion previously made by Sofia and Endal (1980), that changes in the solar luminosity (L) might be accompanied by a change in radius. In order to check the influence of (tiny) solar radius variations on the luminosity, we use the Eddington approximation in Eq. \ref{lumi} which leads to $dL/ L$ = 4$dT/T$ + $dA/A$ (Li et al., 2005), where $T$ is the effective temperature and $A$ is computed through Eq. \ref{surface}.
We are aware that the Sun does not radiate like a black body. If this model is appropriate for the infra-red part of the spectrum or almost true for the visible, by contrast the far UV part departs from it. However, our objective is not to provide a fully comprehensive model of $L(R)$, but to illustrate the effects of observed solar radius variations on global solar parameters such as the luminosity. In this sense this preliminary approximation used is a good indicator: the results obtained may only be illustrative, but are promising. 
It is then straightforward to express $dL/L$, either in the case of an ellipsoidal surface (deriving Eq. \ref{aireellipsoide}, as a function of the parameter $dR$ assuming $dR_{eq}$ = $dR_{pol}$ = $dR$), or in the case of a distorted shape (Eq. \ref{rayonvecteur} with the time-dependent shape coefficients $c_{n}(t)$). Using the Total Solar Irradiance, \textsl{I}, as an indicator of luminosity ($dI/I \propto dL/L$), the modelled irradiance, can now be directly compared to observation data, two parameters being involved: the effective temperature, $T$, and the shape variations, $dR$. We used the irradiance composite dataset updated to October 1, 2003 for which the composite method was established by Fr\"ohlich and Lean (1998)\footnote{Thanks to Fr\"ohlich, C., unpublished data from the VIRGO Experiment on the cooperative ESA/NASA Mission SoHO.}.
We investigated both the ellipsoidal and distorted shape cases.
However, the distorted shape leads to results comparable to the ellipsoid ones (see also Lefebvre and Rozelot,
2003, section 3.2). Hence, we present here only the latter results.
In the case of an ellipsoid, the irradiance temporal variations will be reproduced by a variation $dR$ in the range [10, 200] mas, $dR$ = 0 being the case of a sphere of radius\footnote{Note that $R_{sp}$ is different from the
semi-diameter of the Sun (or standard radius), $R_{\odot}$.} $R_{\odot}$. The choice of the upper limit (200 mas) is given hereafter.
Alternatively, we can adjust the observed $dI/I$ datat to an irradiance model of mean value $I_{0}$, with a temporal sinusoidal variation of period $P$, equal to the solar cycle one, and phase $\phi$:
\begin{equation}
  I_{model} = I_{0} + sin\left( 2\pi t/P + \phi\right) dI .
    \label{irradiance}
\end{equation}
The best fit of the data by $I_{model}$ gives $P$ = 10.09 yrs and $\phi$ = 1.026 rad.
Fig. \ref{irradiancessa} shows the
observed irradiance together with the $I_{model}$ best fit and the
first component ({\emph {RC1}, i.e. the trend)} in the Singular Spectrum Analysis (SSA)
\footnote{Let us recall that the SSA is a technique which has been developed by Vautard et al. (1992). It
has the advantage of working in a data adaptable filter mode instead of
using fixed basis functions, as it is the case for Fourier Transform
or wavelet techniques. Therefore, the SSA has the possibility to get rid of some noise characteristic of a given type of data. 
The SSA is a powerful fast and simple method based on the Principal Component Analysis (PCA) which allows us to filter or reconstruct signals. The basis of the SSA is the eigenvalue-eigenvector decomposition of the
lag-covari\-ance matrix which is composed of the covariances
determined from the shifted time series. Projection of the time
series onto the Empirical Orthogonal Functions (EOFs) yields the
so-called Principal Components (PCs); these are filtered versions
of the original time series. The EOFs are data adaptable to the analogs
of sine and cosine functions while the PCs are the analogs of
coefficients in Fourier analysis.}.
{\emph {RC1}} represents the first
Component in the Reconstruction of the signal. The
{\em RC1} fit is $\chi^2$ = 0.76, better than the sinusoidal
$L_{model}$ fit for which $\chi^2$ = 1.17. Four other curves are
shown: the computed irradiance through Eq. \ref{irradiance} for a
solar ellipsoidal surface (Eq. \ref{aireellipsoide}) with
different ($dR$, $dT$). Computations for an irregular solar shape
(Eqs. \ref{rayonvecteur} and \ref{surface}) lead to similar
results.

\begin{figure}[t]
\begin{center}
\includegraphics[width=7.0cm,height=5.5cm]{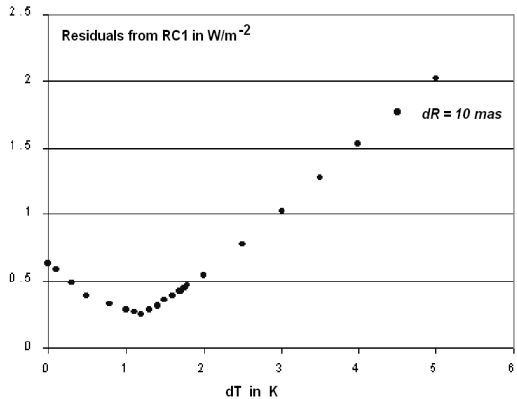}
\end{center}
\caption[]{Computed residuals between the first component {\em RC1} (SSA decomposition) of the observed irradiance and the computed irradiance (see Fig. \ref{irradiancessa}), according to different $dT$. This plot is obtained for $dR$ = 10 mas. The best fit occurs for $dT$ = 1.2 $K$ (other $dR$ leads to larger residuals).
}
\label{irr_dT}
\end{figure}

Computed irradiance is very sensitive to the effective surface
temperature. Two main results appear: (1) Observed irradiance variations can be reproduced with
$dR$ = 200 mas and $dT$ $\approx$ 2$K$, but such a large radius
change is rather unlikely, leaving to no involvement of the magnetic field; (2) an effective surface temperature
variation amplitude $dT$ = 5$K$, whatever $dR$ is, also matches
the observed irradiance variations, but is unlikely too (for the same reason). Hence, in order to quantitatively appreciate the influence of the pair [dR, dT], we computed, inside the limits [0, 200] (mas) for the radius and [0,5] (K) for the temperature, the residuals obtained between the first component in the Singular Spectrum Analysis,
{\em RC1} and our simplified model (for each data point and over nearly two solar cycles). 
A minimum occured when dT = 1.2 K, for dR = 10 mas, as illustrated in Fig. \ref{irr_dT}, which is, among all the figures obtained, that for which the lowest minimum take place (giving thus the best fit; in other words, dR = 10 mas is the {\em lowest minimum} for all dT).

\noindent
A variation of the effective temperature $dT$ = 1.2 $K$ over nearly two solar cycles is close to that obtained by Gray and Livingston (1997) and Caccin et al. (2002) using the ratios of spectral line depths as indicators of the stellar effective temperature. They showed that the solar effective temperature varies systematically during the activity cycle with an amplitude modulation of 1.5 $K$ $\pm$ 0.2 $K$. However, monitoring the spectrum of the quiet atmosphere at the center of the solar disk during thirty years at Kitt Peak, Livingston and Wallace (2003) and Livingston et al. (2005) have shown an immutable basal photosphere temperature within the observational accuracy.

\noindent
We conclude that our fits of modelled irradiance variations (numerical integration through Eqs. \ref{surface} 
and \ref{irradiance}) to observations should be refined.
Thus, we further investigated small solar surface effective
temperature variations ($dT$ $\in$ [0,1.5] $K$) in irradiance
modeling in order to understand the discrepancies between our best
fit $dT$ = 1.2 $K$ at $dR$ = 10 mas, and the latest observations
at Kitt Peak showing $dT$ $\approx$ 0. This yields an unexpected
result. For small values, the phase of irradiance variations with respect to radius ones reverses when crossing the curve plotted in the ($dR$, $dT$)-plane given by
\begin{equation}
    dT_{critical} = 5. 10^{-8} dR^2 +  4.10^{-4} dR  + 0.0005
   \label{phase}
\end{equation}
where $dT$ is in $K$ and $dR$ in mas. This curve distinguishes between correlated (above the $dT_{critical}$ curve) and anticorrelated (below the $dT_{critical}$ curve) solar radius variations with irradiance variations. Consequently, a precise knowledge of $dT$ over the solar cycle is crucial.

In this section, we used the interval $dR$ $\in$ [0, 200] mas to model variations of the irradiance. The lower bound corresponds to a spherical Sun and the upper bound to the value necessary to model all the irradiance variations with only solar radius variations. Those two bounds are unrealistic cases. With respect to the latter interval, $dT_{critical}$ belongs to [0, 0.082]. Hence, we understand the sensitivity of irradiance modeling to very small temperature variations. For example, if observations show that $dT$ $\approx$ 0 with sufficiently small error bars, the Sun is in a state where its radius variations are anticorrelated with irradiance variations (below the $dT_{critical}$ curve).
Since, observations do show that irradiance variations are correlated with the solar activity cycle, we can conclude that solar radius variations are anticorrelated with the solar cycle within the framework of the assumption $dT$ $\approx$ 0 (or, in any case, dT is lower than 0.082 $K$).

Note that the solar subsurface is organized in thin layers (Godier \& Rozelot, 2001) and that changes in these layers have been explored through helioseismology $f$-mode frequencies over the last 9 years. Indeed, Lefebvre \& Kosovishev (2005) and Lefebvre et al. (2007) report a variability of the ``helioseismic" radius in
antiphase with solar activity, the strongest variations
of the stratification being just below the surface (around 0.995
$R_{\odot}$, the so-called {\it ``leptocline''} (Bedding et al. 2007)) while 
the radius of the deepest layers (between 0.97 and 0.99 $R_{\odot}$) change in phase with
11-year activity cycle. These results are fully compatible with
ours and this leptocline layer certainly deserves further
investigations since it is the seat of important
effects (ionization of Hydrogen and Helium, turbulent pressure, shears, inversion of radial rotation gradient,  ...).

\section{ Apparent solar radius variation measurements }\label{solarradius}

So far, the apparent radius of the Sun has been measured from the Earth by different techniques and from different sites. There is an abundant literature
on the subject, but authors still give conflicting results regarding solar radius variations, both in amplitude and in phase.
The discrepancies may come from the determination of the absolute solar apparent radius from the outer layer of the Sun (limb and photosphere) due to solar
atmospheric phenomena (absorption, emission, scattering...), interstellar environment, Earth atmospheric effects and instrumental errors.
Let us illustrate the state of the art. Considering only data obtained at the 150-foot solar tower of the Mount Wilson Observatory, La Bonte and Howard (1981)
found no significant variation of the solar radius with the solar cycle (which was during its ascending phase) when they analyzed magnetograms (Fe I line at 525.0 nm) obtained
routinely from 1974 to 1981. In contrast, Ulrich and Bertello (1995), with the same method, found that the solar radius varied in
phase with the solar cycle over the investigated period 1982–1994 (descending phase), with an amplitude of about 0.4 arcsec.
This variation could be explained by a 3\% change of the line wing intensities during the solar cycle, assuming an apparent faculae and plage surface
coverage of about 15-35\% near the limb, a rather high percentage as emphasized by Bruls and Solanki (2004).
The latter authors also suggest other mechanisms such as a change in the average temperature structure of the quiet Sun (unlikely, according to Livingston and Wallace, 2003)
or an increase in the intensity profile due to the presence of plage emission (faculae, prominence feet...) near the solar limb, associated with magnetic activity variations during a solar cycle.
It can also be argued that the difference between solar radius measurements may come, as suggested by Kosovichev (2005), from an incorrect reduction of
the apparent radius measurements made at different optical depths which are sensitive to the temperature structure.
A recent re-analysis of the magnetograms over 1974--2003 (Lefebvre et al. 2004b, 2006) shows no evident correlation of solar radius variations with magnetic activity (average error bar of 0.07 arcsec). A similar result was found by Wittmann and Bianda (2000), using a drift-time method at Iza$\tilde{n}$a\footnote{Other radius data from Iza$\tilde{n}$a are availlable, such as astrolabe measurements leading to controversial results, which are discussed elsewhere (Badache-Damiani and Rozelot, 2006).} from 1990 to 2000: measurements do not show long-term variations in excess of about $\pm$ 0.0003 arcsec/yr and do not show a solar cycle dependency
in excess of about $\pm$ 0.05 arcsec.

Regarding space measurements of the solar radius, Kuhn et al. (2004) reported an helioseismic upper bound on solar radius variations of only 7 mas ($\pm$ 4 mas)
from the MDI experiment on board SOHO over 1996--2004. The same authors also deduced  an absolute value of the solar
radius, (6.9574 $\pm$ 0.0011)$\times$$10^8$ $m$ or 959.28 $\pm$ 0.15 arcsec, from the Mercury transit of May 7, 2003, even if the instrument was not designed to perform such an astrometric measurement. This value agrees with that deduced from helioseismology, giving confidence in the latter method.

\noindent
Based upon observations, the conclusion is that the solar radius may vary with time (on yearly and decennial time scales), but with a very weak amplitude, certainly
not exceeding some 10--15 mas. We need additional dedicated solar space-based observations (at least balloon flights) to constrain the phase and the amplitude of radius variations. 
And if such observations can be made, we still need a physical model to explain such solar radius variation observations. We address this latter point in the following section.

\section{ Solar radius and luminosity versus gravitational energy variations }

According to the definition of gravitational energy, $E_{g}$= $-\int (G m/r) dm$ (where $r$ is the radial coordinate and $G$ the gravitation constant),
and assuming hydrostatic equilibrium, a thin shell of radius $dr$ containing a mass $dm$ in equilibrium under gravitational and pressure gradient forces
will be expanded or contracted if any perturbation of these forces occurs. However, energy could be stored through gravitational or magnetic fields,
each of them being able to perturb the equilibrium stellar structure, yielding at the end, changes in shape. A possible mechanism could be the following:
if the central energy source remains constant while the rate of energy emission from the surface varies, there must be a reservoir where energy can be
stored or released, depending on the variable rate of energy transport and through several mechanisms like gravitational or magnetic fields.
(Pap et al. 1998, Emilio et al. 2000).

In order to study the consequences of gravitational energy changes
on solar radius variations, Callebaut et al. (2002) used a
self-consistent approach, assuming either a homogeneous or a
non-homogeneous sphere. They calculated $\Delta R/R$ and $\Delta
L/L$ associated with the energies responsible for the expansion of
the upper layer of the convection zone. We use here the same
formalism for a few percent reminder of the modelling TSI (details
of the computations can be found in the above--mentioned paper),
but we consider an ellipsoidal surface (Eq. \ref{aireellipsoide}).
Let $\alpha$ be the fractional radius ($0$ $<$ $\alpha$ $<$ $1$):
if the layer above $\alpha R$ expands, the expansion is zero at
$\alpha$$R$ and is $\Delta$$ R$ at $R$. The increase in height at
a radial distance $r$ in the layer interval $\left(\alpha R,
R\right)$, with $R$ = $R_{sp}$, is given by
\begin{equation}
    h(r) = \frac{(r-\alpha R)^n \Delta R}{R^n (1-\alpha )^n}
    \label{height}
\end{equation}
where $r$ is the usual radial coordinate and $n$= 1, 2, 3... is
the order of the development. The relative increase in thickness
for an infinitesimally thin layer at $r=R_{sp}$ is
$(dh/dr)_{R_{sp}}=  \frac{n \Delta R}{(1-\alpha ) R}$ .
Considering the ideal gas law, $p= \frac{\rho}{m} k T$, and
polytropic law, $p= K \rho ^\Gamma$ (where $\rho$ is the density;
$k$, the Boltzmann constant; $K$, the polytropic constant, and
$\Gamma$, the polytropic exponent --surely an ideal
state--), the relative change in temperature expressed in terms of
the relative change in radius is
\begin{equation}
    \left(\frac{\Delta T}{T}\right)_{R_{sp}} = -\frac{(\gamma -1) n \Delta R }{(1-\alpha ) R}
    \label{temp.}
\end{equation}
where $\Gamma$ can be replaced by $\gamma$, the ratio of the
specific heats. We now apply the above approach to an ellipsoid
with 
$R_{sp} = \sqrt [3] {R_{eq}^2 R_{pol}}$,
using Eq. \ref{aireellipsoide}, and assuming $dR_{eq}\!$ = $\!dR_{pol}\!$ = $\!dR_{sp}$.
When substituting Eq. \ref{temp.} in Eq. \ref{lumi} (Eddington approximation), we obtain
\newpage
\[
\frac{\Delta L}{L}=-\left[\frac{4n(\gamma   -1)}{1-\alpha} + \frac{\frac{a}{c^2}(2a^2-b^2-ab)+\frac{b}{c^3}(2a^3-b^3-ab^2)\ln(\frac{a+c}{b})}{a+\frac{b^2}{c}\ln(\frac{a+c}{b})} \right]
\]
\begin{equation}
\left. ~~~~~~~~ \times~~
\frac{3b}{2b+a}\frac{\Delta R_{sp}}{R_{sp}}
\right.
\label{exp.}
\end{equation}

We made two computations, one with $n$=1 (monotonic expansion
with radius) and the other one with $n$=2 (non monotonic expansion, as shown in Lefebvre and Kosovichev, 2005), using $\gamma$= 5/3, and $\alpha$ $\approx$ 0.96.

Eq. \ref{exp.} implies that a decrease of $R_{sp}$
corresponds to an increase of $L$; that is solar radius and luminosity variations are anticorrelated.

\begin{table}[h]
        \begin{tabular}{llll}
\hline
          $\Delta$L/L =  0.0011 & &  $\Delta$L/L =  0.00073  \\
\hline
$\Delta$R/R = -1.70 $\times 10^{-5}$   & (n=1), &  $\Delta$R/R = -1.13 $\times 10^{-5}$   & (n=1)  \\
(or $\Delta$R = 11.8 km)              &       & (or $ \Delta$R = 7.86 km)  \\
$\Delta$R/R = -8.38 $\times 10^{-6}$ & (n=2), &  $\Delta$R/R = -5.56 $\times 10^{-6}$ & (n=2)  \\
(or $\Delta$R = 5.83 km)               &       &(or $\Delta$R = 3.87 km)  \\
\hline
        \end{tabular}
\caption{Variations of the solar radius computed in two cases: monotonic (n=1) and non monotonic (n=2)
expansion, and for two mean values of $L_{\odot}$. The sign (-) indicates a shrinking. The case $n=2$ is the most likely. }
    \label{delta(r)}
\end{table}

\noindent
Table \ref{delta(r)} gives the results for two values of $\Delta L / L$ = $\Delta I / I$: the usual adopted value, 0.0011, using TSI composite data from 1987 to 2001 (Dewitte et al. (2005); mean value $L_{\odot}$ = 1366.495 $W/m^2$); and 0.00073, determined through a re-analyzis of the composite TSI data over the period of time 1978--2004 (Fr\"ohlich, 2005; mean value $L_{\odot}$ = 1365.993 $W/m^2$).
For n=2 (the most likely case consistent with recent other results), our absolute estimate of $\Delta R_{sp}$ is smaller than the 8.9 km obtained in the case of a spherical Sun by Callebaut et al. (2002). However our $\Delta R_{sp}/R_{sp}$ agrees with that of Antia (2003), i.e. $\Delta R/R$ = 3$\times$$10^{-6}$, who used $f$-mode frequencies data sets from MDI (from May 1996 to August 2002) to estimate the solar seismic radius with an accuracy of about 0.6 km (see also among other authors, Schou et al., 1997 or Antia, 1998 for such a determination of the solar seismic radius to a high accuracy).

\noindent
Three points result from the analysis of the data. The first concerns the ``helioseismic radius" which does not coincide with the photospheric one, the photospheric estimate always being larger by about 300 km (Brown and Christensen-Dalsgaard, 1998).

\noindent
The second point, directly related to our subject, is the shrinking of the Sun with magnetic activity as pointed out
by Dziembowski et al. (2001), using $f$-mode data from the MDI instrument on board SOHO, from May 1996 to June 2000.
They found a contraction of the Sun's outer layers during the rising phase of the solar cycle and inferred a total
shrinkage of no more than 18 km. Using a larger data base of 8 years and the same technique, Antia and Basu (2004) set an upper limit of about 1 km on possible radius variations (using data sets from MDI, covering the period of May 1996 to March 2004). However, they demonstrated that the use of $f$-modes frequencies for $l$ $<$ 120 seems unreliable. 

\noindent
Finally, the third point concerns the luminosity production mechanism, through the parameter {\bf {\textit w}}, called the asphericity-luminosity
parameter. This parameter is defined as
\begin{equation}
  \mbox{\bf {\textit w}} = (dR/R)/(dL/L) .
    \label{asph.lum.}
\end{equation}
According to small observed values of $dR$, a small {\bf {\textit w}} means that $L$ is produced in the upper--most layers (Gough, 2001),
whereas a large {\bf {\textit w}} would imply luminosity production in layers deeper inside the Sun.
>From the above computations and Eq. \ref{asph.lum.}, we can estimate {\bf {\textit w}} as\\
\[ \mbox{\bf {\textit w}}  =  -1.55~10^{-2} ~~~~ (n=1)  ~~~~ \mbox{\rm and  }~~~~  \mbox{\bf {\textit w}} =  -7.61~10^{-3} ~~~~  (n=2)\]
These values\footnote{The sign of $w$ is obviously relevant; it
seems that some authors quoted here have given absolute values.}
(the second is the more likely) can be compared to the ones
computed by Sofia and Endal (1980), -7.5~$10^{-2}$; Dearborn and
Blake (1980), 5.0~$10^{-3}$; Spruit (1992), 2.0~$10^{-3}$; Gough
(2001), 2.0~$10^{-3}$ if the origin of luminosity variations is
located in surface layers, or 1.0~$10^{-1}$ if they are more
deeply seated; and finally to the lower limit given by Lefebvre
and Rozelot (2004), -7.5~$10^{-2}$.

\section{Solar radius variation versus magnetic activity }

As suggested by Livingston et al. (2005), magnetic flux tubes pass between solar granules without interacting with them. Due to magnetic pressure, one could expect a change in the mean size of granules that would be shifted toward the smaller sizes as magnetic activity increases.

\noindent
Such features were confirmed by observations made by Hanslmeier and Muller (2002) at the Pic du Midi Observatory, using the 50-cm refractor (images taken on August 28, 1985 and September 20, 1988).

\noindent
As a consequence, if the number of granules per unit area is constant, the whole size of the Sun would decrease.
This means solar radius variations are anticorrelated with solar magnetic activity.

\section{Conclusions}\label{discussion}

In this study, using a preliminary black-body radiation model for the Sun, we have shown that temporal radius variations must be taken into account in the present efforts to model solar irradiance (we do not claim that irradiance variability is due to radius variability alone). Distortions with respect to sphericity, albeit faint, are related to variations of solar gravitational energy, of surface effective temperature and to variations of luminosity (as solar irradiance is an indicator of solar luminosity). Even if a major simplification was made (using a preliminary black-body radiation model, neglecting magnetic fields which can influence the limb extension), we have obtained constraints on radius and temperature variations through fits to observed irradiance data. Our best fit gives $dT$ = 1.2 $K$ at $dR$ = 10 mas. This surface effective temperature variation agrees with that found by Gray and Livingston (1997) or Caccin et al. (2002). Recent results of Livingston et !
 al. (2005) support a more immutable atmosphere ($dT$ $\approx$ 0). But we have shown that irradiance variation
modelling is very sensitive to small surface effective temperature variation (between 0 and 0.085 K). Indeed, we underlined a phase-shift in the ($dR$, $dT$)--parameter plane between correlated or anticorrelated radius versus irradiance variations.
Better observations of $dT$ might be crucial to determine the phase of radius variations (especially near the limb) with respect to solar cycle activity, noting that observed irradiance variations are in phase with the solar cycle.

We further obtained an upper limit on the amplitude of $dR$, i.e. 3.87 -- 5.83 km, by applying Callebaut's method but taking into account the ellipsoidal shape of the Sun, in a non-monotonic expansion of the radius with depth (in the sub-surface), and composite Total Solar Irradiance.
Our estimate of dR is substantially smaller than the estimate obtained by Callebaut et al. (2002) for a spherical Sun, but it agrees with those derived from helioseismology.

Equating the decrease of radiated energy with the increase of gravitational energy corresponding to the expansion of the upper layer of the convection zone leads to solar radius variations anticorrelated with luminosity ones.
\newline
An estimate of the asphericity-luminosity parameter (\textbf{\textit{w}} = - 7.61 $10^{-3}$)
supports this upper layer mechanism as the source of luminosity variations.

Finally, assuming a constant numbers of granules per unit area, we suggest that solar radius variations might be associated with variations of magnetic pressure between the granules. A possible mechanism could be as follows: as magnetic activity increases, magnetic flux tubes which do not interact with solar granules at the near surface, force the latter to decrease in size; the whole Sun shrinks and radius variations are
thus anticorrelated with solar activity.

The present study was conducted on a large time scale (two solar cycles), and the question of smaller temporal variations (minutes, hours) is not considered here. The above mentioned mechanism may act at a smaller time scale too, but it needs to be confirmed. Space--dedicated missions might be able to answer this question.

\bigskip
\noindent

{\small {\textbf {Acknowledgements.}
Z. Fazel is partly supported by a grant from the French Ministry of Foreign Affairs and the Ministry of Science, Research and Technology (Iran). S. Pireaux acknowledges a CNES post-doctoral grant.}

{\small The authors cordially thank the referees for their remarks which have been used in this version of the paper.}}



\vfill
\end{document}